# Designing Efficient Pair-Trading Strategies Using Cointegration for the Indian Stock Market

Jaydip Sen
*Dept of Data Science*
*Praxis Business School*
Kolkata, INDIA
email: jaydip.sen@acm.org

*Abstract*— **Pair-trading strategy is an approach that utilizes the fluctuations between prices of a pair of stocks in a short-term time frame, while in the long-term the pair may exhibit a strong association and co-movement pattern. When the prices of the stocks exhibit significant divergence, the shares of the stock that gains at price are sold (a short strategy) while the shares of the other stock whose price falls are bought (a long strategy). This paper presents a cointegration-based approach that identifies stocks listed in the five sectors of the National Stock Exchange (NSE) of India for designing efficient pair-trading portfolios. Based on the stock prices from Jan 1, 2018, to Dec 31, 2020, the cointegrated stocks are identified and the pairs are formed. The pair-trading portfolios are evaluated on their annual returns for the year 2021. The results show that the pairs of stocks from the auto and the realty sectors, in general, yielded the highest returns among the five sectors studied in the work. However, two among the five pairs from the information technology (IT) sector are found to have yielded negative returns.**

## I. INTRODUCTION

Pair-trading strategy is a portfolio design approach that exploits the fluctuations between prices of a pair of stocks in a short-term time frame, while in the long-term the pair exhibits a co-movement pattern and mean-eversion behavior. Explained in simple terms, pair-trading identifies a pair of stocks that exhibits a strong correlation over a long-term period while in the short-term there may be a significant variation in their prices. When their prices diverge widening their spread, the stock gaining at price is supposed to be executed with a short strategy while the one experiencing a fall is executed with a long strategy. Here, the term *short* refers to the selling of the shares of stock while the term *long* refers to buying. When the prices of the two stocks exhibit a stable equilibrium, their *spread* reverts to its historical mean value. At that point, the investors can earn profit by reversing their portfolio positions. Pair-trading is considered a very simple yet effective approach to portfolio design. However, based on its simple principles, a set of more complex and powerful approaches to portfolio design and risk mitigation have evolved over the years. In this regard, the principles of long - and short-term return anomalies including lead-lag anomalies, and anomalies of return reversals are worth mentioning [1].

This work proposes a cointegration-based pair-trading approach that identifies stock pairs that exhibit robust cointegration in their prices over three years. Once the cointegrated pairs are identified, the pair-trading portfolios are formed and the performance of the portfolio is observed over a test period of one year. Suitable trigger points are identified using a very robust spread detection mechanism so that the short and the long positions for both stocks are detected accurately. The performance of a portfolio is evaluated based on its annual return. Five sectors are chosen from the National Stock Exchange(NSE) of India. Based on the NSE's monthly report of Dec 31, 2021, the top ten stocks of each of the five sectors are selected [2]. Pair trading portfolios are designed based on pairs from each sector that exhibited cointegration of the *close* prices for the training period Jan 1, 2018, to Dec 31, 2020. The portfolios are tested on the return for the test period Jan 1, 2021, to Dec 31, 2021.

The work has three unique contributions. First, the work proposes a cointegration-based pair trading approach for stock portfolio design that can be used to earn profit by the investors in the stock market. Second, the pair-trading models are trained and evaluated on real-world stock market data and the results are presented to demonstrate the effectiveness of the models. Finally, since the stocks used in the pair-trading portfolios are chosen from several sectors of the NSE, the results are good indicators of the potential profit the investors can earn if they choose to invest in those sectors.

The paper is structured as follows. Section II briefly discusses some related works on pair-trading portfolios. Section III presents an outline of the methodology followed in the research. Section IV presents the results and their analysis. Section V concludes the paper and identifies future work.

## II. RELATED WORK

Portfolio design and optimization is a challenging problem for which numerous solutions and approaches have been proposed by researchers. Statistical arbitrage pair trading refers to a broad approach that involves multivariate pair-trading, generalized pair trading, or statistical arbitrage that has gained popularity over the years [3]. Several approaches to pair-trading-based portfolio design exist in the literature including distance-based approaches [4], cointegration [5], time series [6], and stochastic optimization methods [7]. For efficient portfolio design, various machine learning models have been used for the precise prediction of stock prices [8-10]. Models built on multiobjective optimization [11], principal component analysis [12], deep learning [13-14], and reinforcement learning [15] have also found applications in portfolio optimization and pair-trading.

The present work involves designing portfolios based on cointegrated pairs of stocks from five chosen sectors of the NSE, India. First, from the five sectors, the top ten most influential stocks are identified. All the stocks belonging to a sector are analyzed on their correlation and cointegration characteristics in the training data. Those pairs that are found to exhibit robust cointegration are selected pair-trading. An innovative trading strategy is then designed that provides the investors with reliable signals for triggering the *short*, *long*, or *no action needed* for carrying out transactions in the portfolio for one year from during the portfolio test period. At the end of the test period, all the pairs are evaluated based on the annual return yielded by them. The returns yielded by the pairs also provide an insight into the sectoral profitability for the five sectors studied in this work. In summary, this work while proposing a robust approach to detect cointegration among the stocks of five sectors in the Indian stock market, also presents some useful information on the current returns and profitability of those sectors.

### III. DATA AND METHODOLOGY

The details of the data used and the methodology followed in this work are presented in this section. Cointegrated pairs of stock are first identified based on the training data from five sectors of NSE, and portfolios are built on those pairs. The annual return yielded by the pair-trading portfolios are computed. The pairs and the sectors are finally compared based on the returns of the portfolios over the test period. In the following, the nine steps involved in the research methodology are discussed in detail.

*A. Choice of the Sectors for Analysis*

From the list of sectors listed in the NSE, five are chosen so that they exhibit diversity in the stock market. The chosen sectors are the following: *auto*, *banking*, *information technology* (IT), *pharma,* and *realty*. For each sector, the top ten stocks are selected based on their influence on the respective sectors. The NSE's report published on Dec 31, 2021, is used for selecting the stocks [2].

*B. Extraction of Historical Stock Prices from the Web*

The *DataReader* function of Python's *pandas_datareader* module is used to extract the historical prices of the stocks from the Yahoo Finance website for the period from Jan 1, 2018, to Dec 31, 2021. The *close* values of the stock prices are used for building the pair-trading models. The evaluation is done based on the annual returns of the models on the test data from Jan 1, 2021, to Dec 31, 2021. In the current study, we have chosen pairs from the specific sectors and have not considered pairs from diversified sectors like NIFTY 50.

*C. Computation of the Correlation Matrix of Stock Returns*

At this step, a correlation matrix is generated for each sector using the *heatmap* function of the *seaborn* library to get a rich visualization of the correlation coefficients among the pairs of stocks in that sector. Since there are 10 stocks in a sector, Pearson's product-moment correlation coefficients will be available for 45 distinct pairs. As a convention, the correlation coefficients are computed based on the return values of the *close* prices. For computing the return series for each stock, the Python function *pct_change* function is used. While the correlation coefficients serve as rough indicators of the strength of linear association between a pair of stocks, they are incapable of identifying the presence of any cointegration between the pairs.

*D. Identifying the Cointegrated Pairs of Stocks*

A pair of time series variables are said to be *cointegrated* if they exhibit significant correlation in a long-term time frame while no relationship may be apparent in the short term. To identify the pairs which are cointegrated, the *coint* function defined in the *stattools* submodule under the *statsmodels* module of Python is used. The null hypothesis of the *coint test* assumes that a given pair, which is passed as the two parameters to the function, is not cointegrated [16]. Hence, pairs with $p$-values below the threshold of 0.05 are assumed to be cointegrated. The $p$-values for the 45 pairs of stocks for a given sector are represented in the form of a *heatmap* for effective visualization. The pairs for which $p$-values are less than 0.05 are considered for building the pair-trading portfolios.

*E. Building OLS Regression Model with the Pairs*

The pairs for which the null hypothesis of the cointegration test has been rejected (i.e., the pairs which exhibit cointegration) in Step *D*, an *ordinary least square* (OLS) regression model is built. As a convention, the stock that has a higher mean value of its *close* price is chosen as the predictor (i.e., independent) variable in the OLS model. The OLS function defined in the *linear_model* submodule of the *stasmodels* module of Python is used for building the model. The output of the OLS function has several components. Some important among them are the following. (i) The hedge ratio and the $p$-value of its $t$-statistics, (ii) The $p$-value of the $F$-statistics, (iii) The prob. of the Omnibus test statistics, (iv) The value of the Durbin-Watson test statistics, (v) The prob. of the Jarque-Bera test statistics. The hedge ratio is the coefficient parameter and for the predictor to have significant explanatory power on the target, the $p$-value of its $t$-statistics must be significant. The $p$-value of the $F$-statistic of the model should be significant (i.e., smaller than 0.05) for the regression model to be accurate in its prediction. The Omnibus test statistic should also have a significant $p$-value so that it is confirmed that the predictor has a significant influence on the target. The value of the Durbin-Watson test statistic should be close to 2 so that there is no autocorrelation among the residuals. While values between 0 and 2 indicate the presence of a positive autocorrelation, the residuals are assumed to exhibit a negative autocorrelation if the test statistic lies between 2 and 4. The cointegrated pair of stocks should exhibit significant $p$-values for $F$-statistics and the Omnibus test, and no significant autocorrelation among the residuals of the model. However, there is an additional requirement for cointegration, the residuals should be stationary. This condition is checked in the next step.

*F. Stationarity Analysis of the Residuals of the OLS Model*

As the second step toward validating the cointegration between a pair, the residuals of the OLS model are analyzed for the existence of a unit root (or the non-stationarity of the residual series). The Augmented Dickey Fuller (ADF) test is used for checking the presence of a unit root in the residual series. The *adfuller* function defined in *stattools* submodule of the *statsmodels* module of Python is used for carrying out the ADF test. The ADF test yields test statistics and critical values at 1% and 5% levels of significance. If the test statistic is negative and it is smaller than the critical value at the 1% level of significance, or if the test statistic is positive and it is greater than the critical value at the 1% level of significance, the null hypothesis is rejected. The null hypothesis of the ADF test assumes that there is a unit root in the time series, and hence the series is not stationary. If the null hypothesis fails to gather enough support as judged from its $p$-value, the null hypothesis is rejected, and the series is assumed to be stationary. The ADF test is used to check whether the residual series of the OLS regression model is stationary. The stationarity of the residuals is a requirement for the cointegration of the pair. Hence, the stationarity of the residuals of the OLS is checked.

*G. Generating Trading Signals for the Pairs*

After the cointegration for the pairs is validated, this step involves the generation of the trading signals for the pair. This involves several sub-steps. First, a variable called *ratio* is designed that stores the daily value of the ratio of the *close* value of the predictor time series (also called *asset1*) to the *close* value of the target (also called *asset2*). Second, the standardized values of the variable ratio are computed by subtracting the mean value from each ratio value and dividing the result by the standard deviation of the ratio values. The upper and lower limits of the Z-scores of the ratio values are computed by adding and subtracting the standard deviation from the mean value, respectively. Since, the mean value of the Z-scores is 0 and their standard deviation is 1, hence the values of the upper and lower limit of the Z-score of the ratio are 1 and -1, respectively. Third, a *pandas* dataframe called *signals* is now created with the following six columns: (i) *date* (ii) *asset1*, consisting of the time series of the *close* values of the first stock, (iii) *asset2*, consisting of the time series of the *close* values of the second stock, (iv) *Z score*, consisting of the standardized values of the ratio, (v) *upper limit*, storing the upper limit of the *Z* scores (all entries are 1 under this column), and (vi) *lower limit*, storing the lower limit of the *Z* scores (all entries are -1 under this column).

Based on the *Z*-scores and their upper and lower limits, the trading signals for *asset1* are now constructed. If the *Z*-score at a point exceeds the *upper limit*, then *asset1* should be used with the *short* strategy, else the *long* strategy should be followed. The *short* strategy for *asset1* involves selling *asset1*, while the "long" strategy of *asset1* involves buying more units of *asset1*. A column *signals1* is now added to the *signals* dataframe with an entry -1 for the *short* cases and entry of 1 for the *long* cases for *asset1*. Another column *signals2* is also added to the *signals* dataframe for *asset2*. The entries under *signals2* differ in sign with those under *signals1*. In other words, records with entry 1 under *signals1* have an entry -1 under *signals2* and vice versa. This is because the trading strategies for the two assets are complementary in nature. At this point, the *signals* dataframe consists of the following columns: (i) *date*, (ii) *asset1*, (ii) *asset2*, (iii) *Z*-score, (iv) *upper limit*, (v) *lower limit*, (vi) *signals1*, and (vii) *signals2*.

*H. Identifying the Points of Investment Opportunities*

To identify the portfolio *positions*, i.e., the points of undervalued investment opportunities for the two stocks, two additional columns are added to the *signals* dataframe. For the first stock (i.e., *asset1*), column *positions1* is added. The entries under this column are computed based on the first-order difference of the *singals1* column. Similarly, for the second stock, column *positions2* is added based on the first-order difference values of the *signals*2 column. An entry of 0 under *positions1* implies that there is no change in the current value of *signals1* with its previous value. An entry of 1 indicates that *signals1* has exhibited a change from 0 to 1, while an entry of -1 implies that the change is from 1 to 0. Similarly, the entries under column *positions2* are determined. The records with *positions1* value 1 correspond to a *long* trigger for *asset1*, while *positions1* values 0 indicate a *short* trigger for *asset1*. A record with positions1 entry of 1 indicates that *signal1* has changed its value from 0 to 1. In other words, *asset1* has received a *long* trigger. Similarly, a *short* trigger is exhibited by a *positions1* entry of -1. An entry of 0 for *postions1* indicates no action is needed for *asset1*. Similarly, entries of -1, 1, and 0 for *positions2* indicate a *short* trigger, a *long* trigger, or no action needed, respectively for *asset2*. A graph is plotted to identify the *short* and *long* trigger points for both stocks during the portfolio testing phase (i.e., over the year 2021).

*I. Computation of the Returns of the Pair Portfolio*

In the final step, the portfolio returns for the year 2021 is computed. The pair-trading portfolio is formed on Jan 1, 2021, with an initial capital of 100000 units for each asset (i.e., with a total investment of 200000 units). For each day over the test period, the sum of the holding values and the cash amount for both the stocks are computed and stored in a variable *total portfolio value*. At the end of Dec 31, 2021, the excess (or deficit) of the total portfolio value over the total initial investment of 200000 units is computed to arrive at the profit (or loss). In this computation, the transaction costs are ignored. Based on the computed profit, the return of the portfolio is derived.

## IV. EXPERIMENTAL RESULTS

This section presents the detailed results of the performances of the pair-trading portfolios for different sectors. Python 3.9.7 and its associated libraries of numpy, pandas, matplotlib, statsmodels, and seaborn are used for data extraction, designing, and testing of the pair-trading models. The models are trained and validated on the GPU environment of Google Colab.

*A. Auto Sector*

Ten pairs in this sector yield significant *p*-values for their cointegration, as depicted in Fig 1. Among these ten pairs, the following six are studied in this work: (i) Bharat Forge (BF) and Ashok Leyland (AL), (ii) Eicher Motors (EM) and Ashok Leyland (AL), (iii) Maruti Suzuki (MS) and Eicher Motors (EM), (iv) Maruti Suzuki (MS) and Ashok Leyland (AL), (v) Eicher Motors (EM) and Bharat Forge (BF), and (vi) Maruti Suzuki (MS) and Bharat Forge (BF).

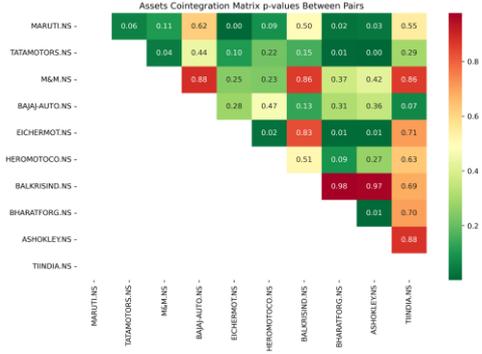

Fig 1. The *p*-values for the cointegration between pairs for the auto sector

Due to space constraints, the results of the pair BF and AL are presented in detail. For the other pairs, only the annual return figures for the year 2021 are presented. The daily *close* prices of BF and AL are depicted for the training period (i.e., Jan 1, 2018, to Dec 31, 2020) in Fig 2 to visually check the long-term association between them. As discussed in Section III, the next step involves building an *ordinary least square* (OLS) regression model with the BF time series as the *predictor* variable and the AL time series as the *target*. The regression results are presented in Fig 3. The *hedge ratio* (i.e., the coefficient parameter for BF) is found to be 0.1839 with a standard error of 0.01. The highly significant *p*-value of the *F*- statistic indicates that there is a strong linear relationship between the two series. The Jarque-Bera test yielded a highly significant *p*-value for the test statistic indicating the skewness and the kurtosis of the residuals of the model are not like those of a normal distribution. The Durbin-Watson test statistic lies between 0 and 2 indicating a *positive autocorrelation* among the residual. Finally, the Omnibus test results support the *p*-value of the *F* statistic indicating the presence of a strong linear relationship between the *target* and the *predictor*. In the next step, the residuals of the model are computed and plotted.

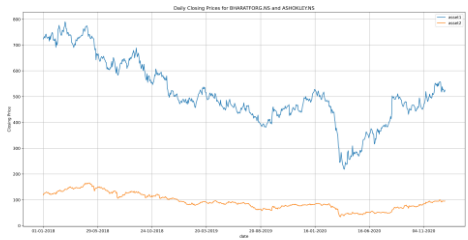

Fig 2. Daily *close* prices plot for Bharat Forge and Ashok Leyland

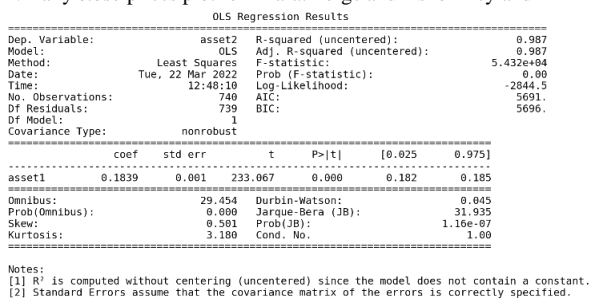

Fig 3. The results of the OLS regression model with Ashok Leyland (asset2) as the target variable and Bharat Forge (asset1) as the predictor variable

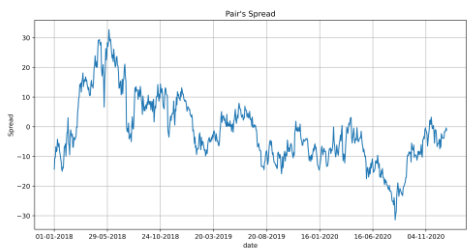

Fig 4. The residual plot of the BF- AL OLS model

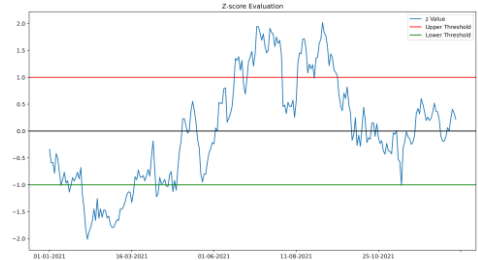

Fig 5. The Z-values of the ratio series and their limits for BF and AL

Fig 4 depicts the residual plot which looks like a stationary series. The Augmented Dickey Fuller (ADF) test on the residual series yielded a value of -2.7397 while the critical value at 1% level of significance is -3.4392. The results of the ADF test indicated that the residual series is not stationary. However, as the BF and AL series are cointegrated, we proceed to execute with their pair-trading strategies. Fig 5 depicts the plot of the Z-values of the *ratio* series and their upper and lower limits. The *ratio* series represents the ratio of the daily *close* prices of BF and AL for the training data. The pair-trading signal points are plotted in Fig 6, in which four types of signal points are identified, *long* strategy for BF, *short* strategy for BF, *long* strategy for AL, and *short* strategy for AL. The four strategies are denoted by different symbols in Fig 6. Finally, with an initial investment of 1000000 units of capital for each of the two stocks, the trading strategy is initiated on Jan 1, 2021. Fig 7 depicts the daily values of the portfolio for the year 2021. The value of the portfolio at the end of the year is found to be 235269 which corresponds to an annual return of 17.63%. Table I presents the annual returns for the six pairs of stocks from the auto sector which exhibited significant *p*-values in the cointegration test. All the pairs are found to produce positive returns with the BF-AL pair producing the highest return.

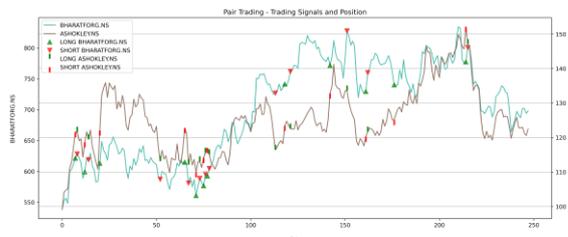

Fig 6. The pair trading scenario for the stocks BF and AL – trading signals and their positions identified

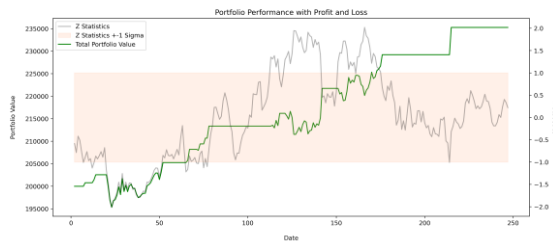

Fig 7. The total value of the pair trading portfolio of BF and AL for the year 2021

TABLE I. ANNUAL RETURNS OF AUTO SECTOR PAIR TRADING

| Stock Pair | Init Investment | Profit | Annual Return |
|---|---|---|---|
| BF - AL | 200000 | 35270 | 17.63 |
| EM - AL | 200000 | 27773 | 13.89 |
| MS - EM | 200000 | 23968 | 11.98 |
| MS - AL | 200000 | 22503 | 11.25 |
| EM - BF | 200000 | 21608 | 10.80 |
| MS - BF | 200000 | 20300 | 10.15 |

*B. Banking Sector*

There are only three pairs from this sector that are found to have significant *p*-values for their cointegration as depicted in Fig 8. However, two additional pairs are included for the analysis as their *p*-values are also close to the threshold value of 0.05. The five pairs which are studied under this sector are as follows: (i) State Bank of India (SB) and IDFC First Bank (IF), (ii) Federal Bank (FB) and IDFC First Bank (IF), (iii) HDFC Bank (HD) and Kotak Mahindra Bank (KM), (iv) ICICI Bank (IC) and Kotak Mahindra Bank (KM), and (v) Axis Bank (AX) and State Bank of India (SB).

While the results of the pair SB and IF are discussed in detail, for the remaining four pairs, only their annual return values for the year 2021 are presented. The daily *close* prices of SB and IF are plotted in Fig 9. The OLS regression results with IF as the target variable and SB as the predictor variable are presented in Fig 10. The hedge ratio is 0.1478, and the *F*-statistic has a highly significant *p*-value. The residual series is plotted in Fig 11.

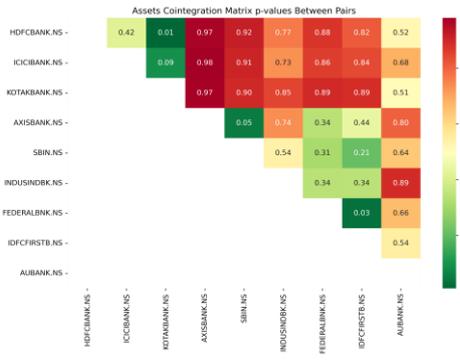

Fig 8. The *p*-values for the cointegration between pairs for the banking sector

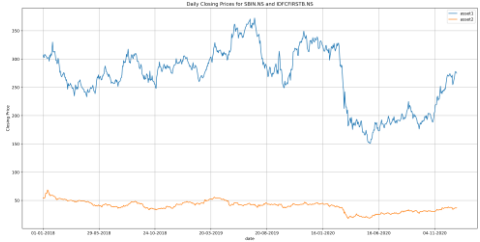

Fig 9. Daily close prices plot for SBI and IDFC First Bank

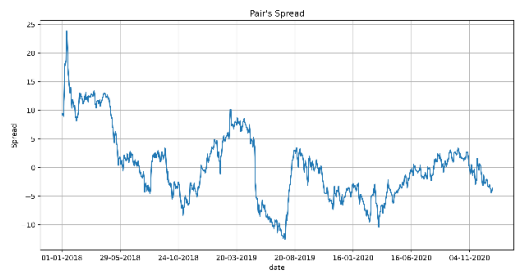

Fig 10. The results of the OLS regression model with IDFC First Bank (*asset2*) as the target and SBI (*asset1*) as the predictor

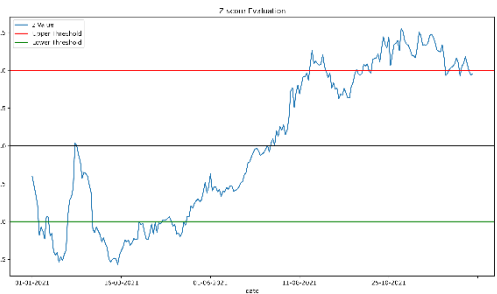

Fig 11. The residual plot for the SB-IF OLS model

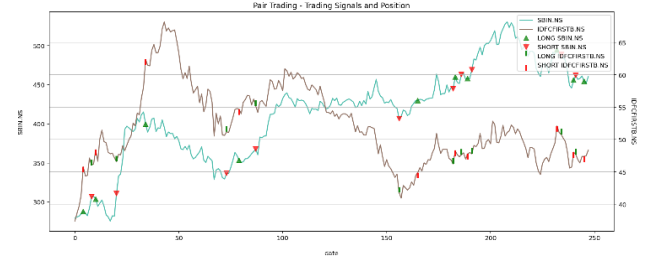

Fig 12. The Z-values of the ratio series and their limits for SB and IF

The ADF test on the residual series yields a test statistic value of -2.5566 while the critical value at 1% level of significance is -3.4392. Hence, the residual series is not stationary. However, the pair trading strategy is still executed.

The plot of the Z-values of the *ratio* series and their upper and lower limits are shown in Fig 12. Fig 13 and Fig 14 show the trading signal points, and the daily total value of the portfolio for 2021, respectively. With an initial investment of 200000 on Jan 1, 2021, the total value of the portfolio at the end of 2021 is found to be 219926. The return yielded by the portfolio is 9.96%. Table II presents the returns of the portfolios of the five pairs of stocks in the banking sector, in which except for the pair AX-SB, all other pairs have produced positive returns.

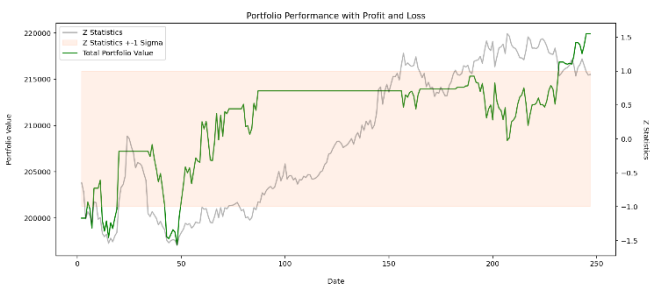

Fig 13. The pair trading scenario for the stocks SB and IF – trading signals and their positions identified

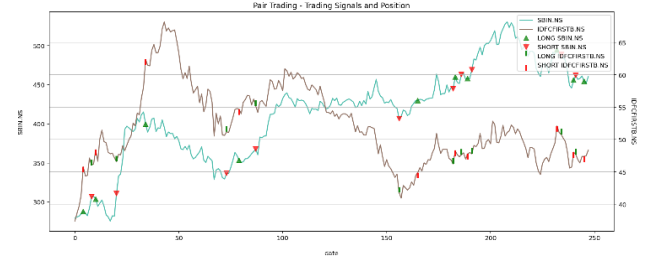

Fig 14. The total value of the pair trading portfolio of SB and IF for the year 2021

TABLE II. ANNUAL RETURNS OF BANKING SECTOR PAIR TRADING

| Stock Pair | Init Investment | Profit | Annual Return |
|---|---|---|---|
| SB - IF | 200000 | 19926 | 9.96 |
| FB - IF | 200000 | 19300 | 9.65 |
| HD - KM | 200000 | 11056 | 5.53 |
| IC - KM | 200000 | 3638 | 1.82 |
| AX - SB | 200000 | -17575 | -8.79 |

*C. IT Sector*

As depicted in Fig 15, six pairs exhibited significant *p*-values for their cointegration. These six pairs are as follows: (i) TCS (TC) and Coforge (CF), (ii) Infosys (IF) and HCL Tech (HC), (iii) Tech Mahindra (TM) and L&T Tech Services (LS), (iv) Wipro (WP) and L&T Tech Services (LS), (v) TCS (TC) and Wipro (WP), and (vi) HCL Tech (HC) and L&T Infotech (LI). The results for the pair TC-CF are presented in detail. Fig 16 exhibits the *close* values of the TC and the CF series for the period Jan 1, 2018, to Dec 31, 2020. The results of the OLS regression model built with TC as the predictor and CF as the target are presented in Fig 17. The hedge ratio is 0.7061. The highly significant *p*-value of the *F*-statistics indicates that there is a strong linear relationship between the two series TC and CF.

The residuals of the OLS model are plotted in Fig 18. The Z-values of the ratio with their upper and lower bounds are depicted in Fig 19. The ADF test on the ratio series yielded a test statistic value of -1.2592, while the critical values at 1% level of significance is -3.4392. The ADF test results indicate that the residual series is not stationary. However, the pair trading strategy is executed since the two series were found to be cointegrated.

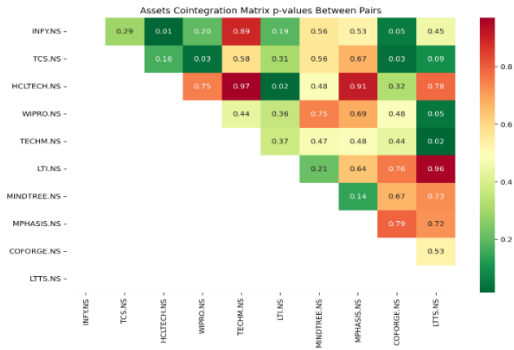

Fig 15. The *p*-values for the cointegration between pairs for the IT sector

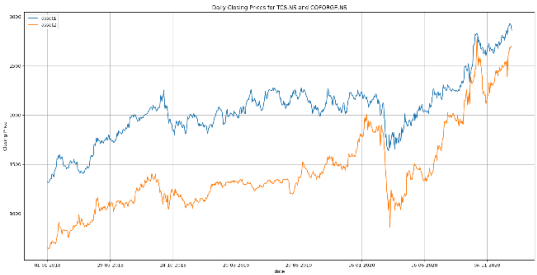

Fig 16. Daily close prices plot for TCS (*asset1*) and Coforge (*asset2*)

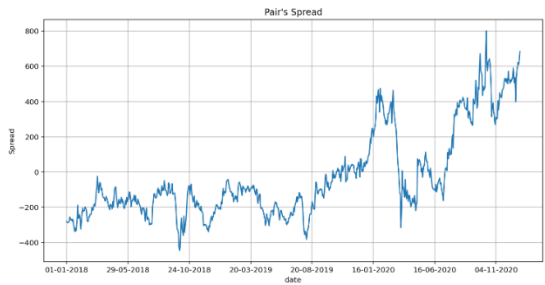

Fig 17. The results of the OLS regression model with Coforge (*asset2*) as the target and TCS (*asset1*) as the predictor

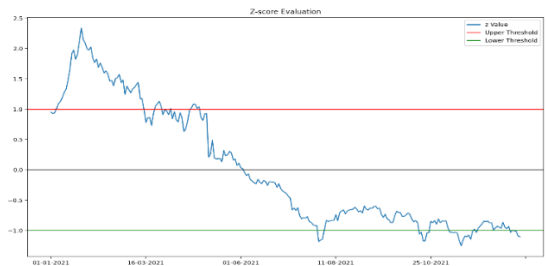

Fig 18. The residual plot for the TC-CF OLS model

Fig 19. The Z-values of the ratio series and their limits for the TC and CF

The signal points of the pair trading during 2021 are depicted in Fig 20. Fig 21 exhibits the daily portfolio value for the pair for the year 2021. The total portfolio value for the TC-CF pair at the end of 2021 is 217462. For an initial investment of 200000, the pair yields a return of 8.73%. Table III presents the portfolio return for the six pairs from the IT sector. Except for two pairs (i.e., TC-WP and HC-LI), the remaining four pairs have produced positive returns.

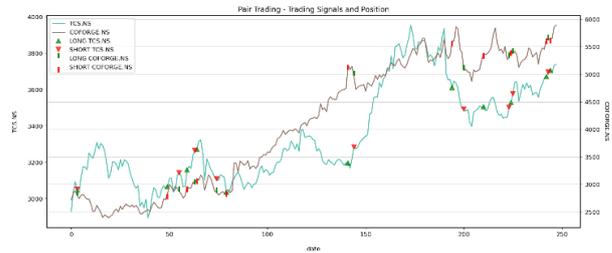

Fig 20. The pair trading scenario for the stocks TC and CF – the trading signals and their positions identified

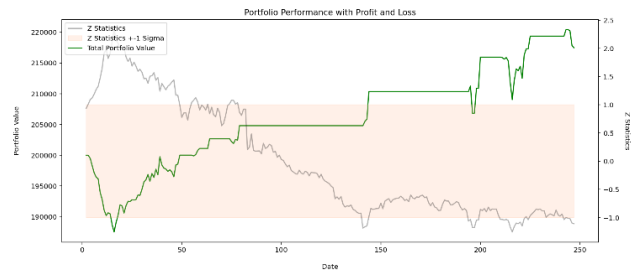

Fig 21. The total value of the pair trading portfolio of TC and CF for the year 2021

TABLE III.  ANNUAL RETURNS OF IT SECTOR PAIR TRADING

| Stock Pair | Init Investment | Profit | Annual Return |
|---|---|---|---|
| TC - CF | 200000 | 17460 | 8.73 |
| IF - HC | 200000 | 10940 | 5.47 |
| TM - LS | 200000 | 6720 | 3.36 |
| WP - LS | 200000 | 3740 | 1.87 |
| TC - WP | 200000 | -8660 | -4.33 |
| HC - LI | 200000 | 13580 | -6.79 |

*D. Pharma Sector*

Fig 22 shows that two pairs have significant *p*-values for their cointegration. Three additional pairs are included in the study as their *p*-values are just marginally higher than the threshold value of 0.05. The five pairs in the pharma sector which are analyzed are as follows: (i) Lupin (LP) and Alkem Labs (AK), (ii) Lupin (LP) and Biocon (BI), (iii) Dr. Reddy's Labs (DR) and Divi's Labs (DV), (iv) Cipla (CI) and Biocon (BI), and (v) Lupin (LP) and Laurus Labs (LR).

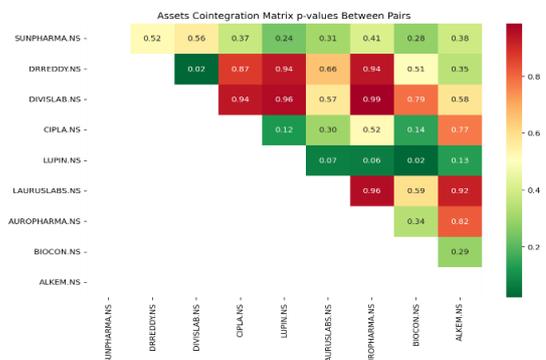

Fig 22. The *p*-values for the cointegration between pairs for the pharma sector

The pair LP-AK is analyzed in detail. The *close* values for the LP and the AK series for the period Jan 1, 2018, to Dec

31, 2021, are plotted in Fig 23. The results of the OLS regression model fitted on the training data with LP as the predictor and AK as the target are exhibited in Fig 24. The hedge ratio is 2.6030. A strong linear relationship between the two series is indicated by a highly significant *p*-value of the *F*-statistic.

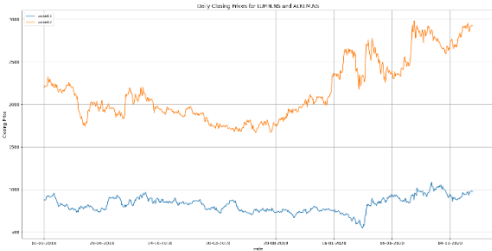

Fig 23. Daily close prices plot for Lupin (*asset1*) and Alkem Labs (*asset2*)

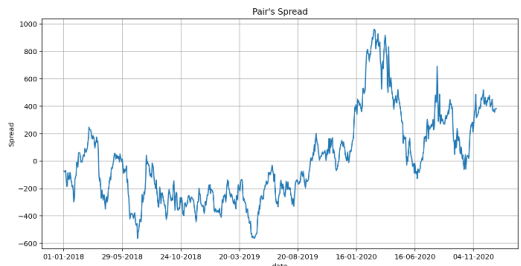

Fig 24. The results of the OLS regression model with Alkem Labs (*asset2*) as the target variable and Lupin (*asset1*) as the predictor variable

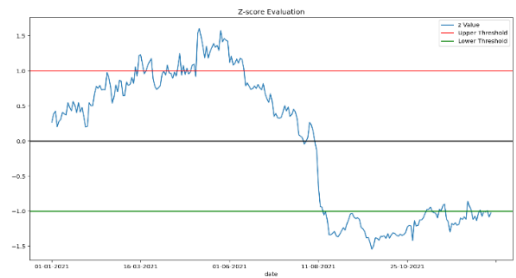

Fig 25. The residual plot for the LP-AK OLS model

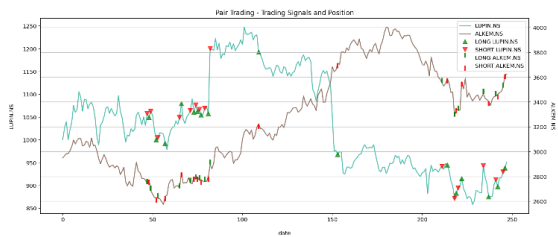

Fig 26. The Z-values of the ratio series and their limits for LP and AK

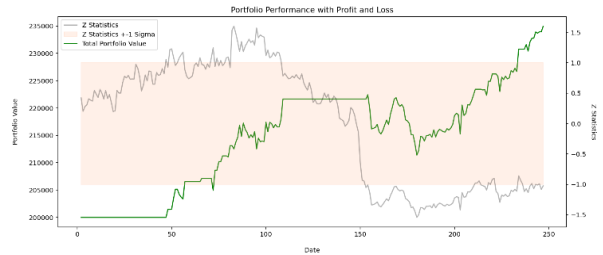

Fig 27. The pair trading scenario for the stocks LP and AK – the trading signals and their positions identified

The residual values are plotted in Fig 25. The Z-values of the ratio and their upper and lower bounds are depicted in Fig 26. The ADF test on the ratio series yielded a test statistic value of -2.0947, while the critical value at a 1% level of significance is -3.4392. The ADF test results indicate that the residual series is not stationary. However, the pair trading strategy is executed as the pair is found to be cointegrated. The signal points for the *short* and *long* transactions for the pair trading strategy in 2021 are presented in Fig 27. The daily portfolio values for the pair in 2021 are exhibited in Fig 28. The portfolio of the LP-AK pair is found to have yielded a profit of 34986 with an initial investment of 200000. Hence, the portfolio produced a return of 17.49%. Table IV presents the returns for the five pairs from the pharma sector. Barring LP-LR, the other pairs have yielded positive returns.

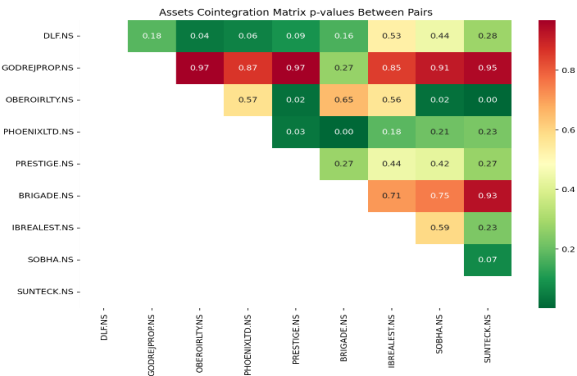

Fig 28. The total value of the pair trading portfolio of LU and AK for the year 2021

TABLE IV. ANNUAL RETURNS OF PHARMA SECTOR PAIR TRADING

| Stock Pair | Init Investment | Profit | Annual Return |
|---|---|---|---|
| LP - AK | 200000 | 34986 | 17.49 |
| LP - BI | 200000 | 26993 | 13.50 |
| DR - DV | 200000 | 10942 | 5.47 |
| CI - BI | 200000 | 7614 | 3.81 |
| LP - LR | 200000 | -15812 | -7.91 |

*E. Realty Sector*

As depicted in Fig 29, six pairs in this sector have significant *p*-values for their cointegration. Alongside these six pairs, another pair is included in the analysis since its *p*-value is also very close to the threshold value of 0.05. The seven pairs included in the analysis are as follows: (i) Oberoi Realty (OR) and Prestige Estate Projects (PE), (ii) DLF (DL) and Oberoi Realty (OR), (iii) Phoenix Mills (PM), and Prestige Estate Projects (PE), (iv) Oberoi Realty (OR) and Sunteck (ST), (v) Oberoi Realty (OR) and Sobha (SB), (vi) Brigade Enterprises (BE) and Godrej Properties (GP), (vii) Phoenix Mills (PM) and Brigade Enterprises (BE).

Fig 29. The *p*-values for the cointegration between pairs for the realty sector

The pair OR-PE is analyzed in detail. The results of the OLS regression model built using the PE series as the target variable and the OR as the predictor are presented in Fig 30. The residual series is plotted in Fig 31. The ADF test produced a test statistic value of -3.4499 while the critical value at 1% level of significance is -3.4392. The ADF test results indicate that the residual series is stationary. The signal points for the OR-PE pair trading in 2021 are presented in Fig 32. Fig 33 exhibits the daily portfolio values for the pair in 2021. The OR-PE pair has yielded a profit of 32488 with an initial investment of 200000. The return for this pair is 16.24. Table V presents the returns of the seven pairs of the realty sector.

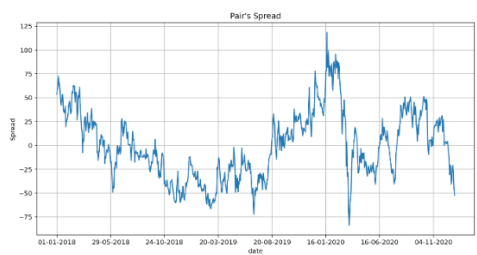

Fig 30. The results of the OLS regression model with Prestige Estate (*asset2*) as the target and Oberoi Realty (*asset1*) as the predictor

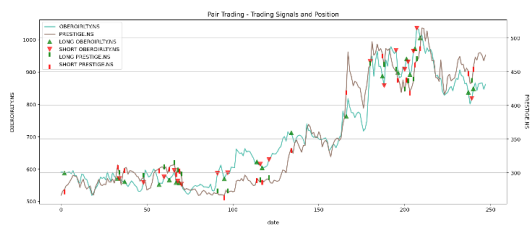

Fig 31. The residual plot for the OR-PE OLS model

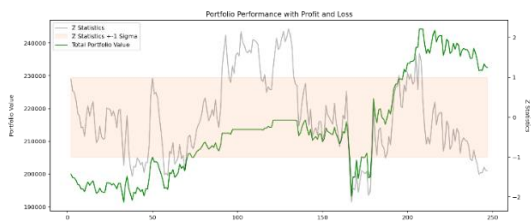

Fig 32. The pair trading scenario for the stocks OR and PE – the trading signals and their positions identified

Fig 33. The total value of the pair trading portfolio of OR and PE for the year 2021

TABLE V. ANNUAL RETURNS OF REALTY SECTOR PAIR TRADING

| Stock Pair | Init Investment | Profit | Annual Return |
|---|---|---|---|
| OR - PE | 200000 | 32488 | 16.24 |
| DL - OR | 200000 | 27337 | 13.67 |
| PM - PE | 200000 | 27184 | 13.59 |
| OR - ST | 200000 | 24481 | 12.24 |
| OR – SB | 200000 | 11711 | 5.86 |
| BE – GP | 200000 | 8339 | 4.17 |
| PM - BE | 200000 | 4457 | 2.23 |

A summary of results is exhibited in Table VI, in which for each sector, the number of pairs studied, the number of pairs that produced positive returns, and the maximum return yielded by a pair for that sector are shown. All pairs from the auto sector and the pharma have produced positive returns. Two pairs of the IT sector have yielded negative returns and this sector also has yielded the lowest value of the max return.

TABLE VII. SUMMARY RESULTS OF PAIR TRADING PORTFOLIOS

| Sector | No of Pairs | Positive Return Pairs | Max Ret |
|---|---|---|---|
| Auto | 6 | 6 | 17.63 |
| Pharma | 5 | 4 | 17.49 |
| Realty | 7 | 7 | 16.24 |
| Banking | 5 | 4 | 9.96 |
| IT | 6 | 4 | 8.73 |

## V. CONCLUSION

This paper has presented a cointegration-based pair-trading approach for stocks chosen from five sectors listed in the NSE of India. The top ten stocks are selected from each sector and pair-trading portfolios are designed with the cointegrated pairs. The pair-trading portfolios are designed on the stock price data from Jan 1, 2018, to Dec 31, 2020, and they are evaluated over the year 2021. The auto sector and the realty sector produced the best results as all the cointegrated pairs from these two sectors yielded positive returns. However, two pairs from the IT sector and one each from the banking and the pharma sectors were found to have negative returns. Comparison of the cointegration-based approaches with the clustering-based approaches is a future research plan.